\documentstyle[12pt,a4,epsfig]{article}

\newcommand{\beq}{\begin{equation}}
\newcommand{\eeq}{\end{equation}}
\newcommand{\beqar}[1]{\begin{eqnarray}\label{#1}}
\newcommand{\eeqar}{\end{eqnarray}}

\newcommand{\al}{\alpha}
\newcommand{\be}{\beta}
\newcommand{\ep}{\varepsilon}

\newcommand{\de}{\delta}

\newcommand{\la}{\lambda}

\newcommand{\om}{\omega}

\newcommand{\as}{\alpha_S}

\def\eq#1{{Eq.~(\ref{#1})}}

\def\npb#1#2#3{    {\it Nucl. Phys. }{\bf B#1} (19#2) #3}
\def\plb#1#2#3{    {\it Phys. Lett. }{\bf B#1} (19#2) #3}
\def\prd#1#2#3{    {\it Phys. Rev. }{\bf D#1} (19#2) #3}

\def\prl#1#2#3{    {\it Phys. Rev. Lett. }{\bf #1} (19#2) #3}

\def\zpc#1#2#3{    {\it Z. Phys. }{\bf C#1} (19#2) #3}

\relax
\begin{document}
\vspace*{-2cm}
\begin{flushright}
 NTZ 8/98\\
TAUP 2488-98\\
DTP-98-16 \\
\end{flushright}
\vspace{2cm}
%%%%%%%%%%%%%%%%%%%%%%%%%%%%%%%%%%%%%%%%%%%%%%%%%%%%%%%%%%%%%%%%
\begin{center}
{\LARGE \bf
 The BFKL Pomeron \\  in 2+1 dimensional  QCD
\footnote{Supported by
German Bundesministerium f\"ur Bildung, Wissenschaft, Forschung und
Technologie, grant No. 05 7LP91 P0, and
by the Volkswagen Stiftung.}}\\[2mm]
\vspace{1cm}
{ \bf D.Yu.~Ivanov$^{a), b)}$, R.~Kirschner$^{a)}$,
E.M.~Levin$^{c),d) }$,
L.N.~Lipatov$^{d)}$,
L.~Szymanowski$^{a), e)}$ and M.~W\"usthoff$^{f)}$}\\
\vspace{1cm}

$^{a)}$Naturwissenschaftlich-Theoretisches Zentrum  \\
und Institut f\"ur Theoretische Physik, Universit\"at Leipzig
\\
Augustusplatz 10, D-04109 Leipzig, Germany
\\
\vspace{2em}
$^{b)}$
Institute of Mathematics, 630090 Novosibirsk, Russia \\

\vspace{2em}
$^{c)}$ School of Physics and Astronomy, Tel Aviv University
\\
Ramat Aviv, 69978 Israel\\
\vspace{2em}
$^{d)}$ St. Petersburg Nuclear Physics Institute \\
188350. Gatchina, St. Petersburg, Russia \\
\vspace{2em}
$^{e)}$
Soltan Institute for Nuclear Studies, Ho\.za 69, 00-681 Warsaw, Poland

\vspace{2em}
$^{f)}$ University of Durham, Dept. of Physics \\
South Rd., Durham City DH1 3LE, United Kingdom

\end{center}
\newpage

\vspace{5cm}
\centerline{\bf Abstract:}
We investigate the high-energy scattering in the spontaneously broken
Yang - Mills gauge theory in 2+1 space--time dimensions and present the
exact solution of the leading $\ln s$ BFKL equation.
The solution is constructed in terms of
special functions using the earlier results of two of us (L.N.L.
and L.S.). The analytic properties of the $t$-channel partial wave as
functions of the angular momentum and momentum transfer have been
studied.  We find in the angular momentum plane:  (i) a Regge pole
whose trajectory  has an intercept larger than 1 and (ii) a fixed cut
with the rightmost singularity located at $j=1$.  The massive
Yang - Mills theory can be considered as a theoretical model for the
(non-perturbative) Pomeron. We study the main structure and property
of the solution including the Pomeron trajectory at
momentum transfer different from zero.  The relation to the results of
M.~Li and C-I. Tan for the massless case is discussed.

\vspace*{\fill}
\eject
\newpage

\section{Introduction}

Recent  experimental data from HERA \cite{HERADATA} on deep inelastic
scattering at small $x$  and fixed $Q^2$
and from the TEVATRON on high-energy diffraction
\cite{Schlein} revived
the interest in the longstanding problem of the Pomeron structure and
of the relation between soft  and hard
processes at high energy.

For the hard Regge processes one can use the BFKL theory
\cite{klf}, but we are lacking a selfconsistent theoretical approach to
the soft Pomeron and have to rely merely on general properties of
analyticity, causality and crossing symmetry in developing an extended
and successful phenomenology of high-energy soft interactions
\cite{DL}, \cite{GLMS}, \cite{ELLEC}.

Some theoretical understanding of the Pomeron has been derived from the study
of the leading $\ln s$ approximation of superrenormalizable models
like $\la \phi^3$ in 3 + 1 dimensions. The main features of the result
have been included in the parton model of peripheral interactions
and they are the basis of our understanding of the Pomeron structure
\cite{FEYN}, \cite{GRI}.  However, such models result in Regge
singularities with intercept around $-1$ and do not reproduce  essential
features of the Pomeron. Much effort has been applied to show the
selfconsistency of the Pomeron hypothesis in the framework of reggeon
field theory or Gribov's Reggeon calculus \cite{GRIB}. A Reggeon field theory
approach to QCD has been developed in \cite{WHITE}.

On the contrary, for the hard Pomeron we can apply perturbative QCD
and derive a number of detailed predictions \cite{PQCDA}. A
special role plays  the BFKL pomeron \cite{klf} appearing in the leading
$\ln s$ ($\approx \ln \frac{1}{x}$) approximation. The
main features of the BFKL Pomeron, however, 
look different from properties of the
soft Pomeron.

In this paper we study the BFKL Pomeron in spontaneously broken 2 + 1
dimensional gauge theory, using previous results obtained in ref.
\cite{preprint}. One can consider this theory as a
simple model for the soft Pomeron.  Indeed we show that the
resulting BFKL Pomeron is a normal moving Regge pole with its intercept
$\al_P(0)>1.$

The coupling of this theory has the dimension of mass. The interaction is
superrenormalizable. This results in the
absence of scaling violations of structure functions
due to ultraviolet divergences.
 On the other
hand the infrared singularities in the massless limit are stronger compared
to 3 + 1 dimensional QCD. The comparison allows us to discuss  the
influence of the ultraviolet and infrared singularities on the Pomeron
structure.

In QCD (massless gluons in 3 + 1 dimensions) the known way \cite{LIP86}
of solving the BFKL equation relies on conformal symmetry. This
approach is useless in the case of massive gauge bosons. Up to now the
solution is not known for the massive case. In the special case
of 2 + 1 dimensions, however, the equation exhibits a 
simple iterative structure which allows to construct a solution. The experience
gained in the 2 + 1 dimensional theory will be helpful in solving the
corresponding equation in the physical case.

We obtain the exact solution both for the forward and non-forward
cases, and calculate the partial wave amplitude for the scattering of
two massive gauge bosons. We investigate the Regge singularities in the complex
angular momentum plane and their behaviour in dependence of the momentum
transfer.

The paper is based on an early investigation by two of the authors
\cite{preprint}, where the basic idea of the iterative solution was
formulated for the general non-forward case.
This investigation was motivated in particular by \cite{marchesini},
where the BFKL equation with the infrared regularization  has
been considered.
We discuss the relation of our result
with the one by M.~Li and C.-I. Tan \cite{TAN} where the massless 2 + 1
dimensional gauge theory has been considered.

\section{BFKL equation with massive gluons }
\subsection{$3+1$ dimensions}
\setcounter{equation}{0}
Let us start with the short reminder
of the results  obtained within
the leading logarithmic approximation of perturbation theory (LLA)
for the amplitudes of the high energy scattering
in the spontaneously broken Yang--Mills theory
 \cite{klf}.
We discuss the simplest case of $SU(2)$ gauge group with symmetry
breaking by one Higgs doublet (fundamental representation).
This is the case discussed in \cite{klf}; we shall follow the notation of
that paper. The generalization to $SU(N)$ gauge group is
straightforward and is done in the section 3.3.  Notice that the
details depend on the way of symmetry breaking. We consider the case
that all gauge bosons become massive.

The amplitude describing the elastic
two-particle scattering $AB \to A^\prime B^\prime$
can be decomposed into the amplitudes with definite
isotopic spin $T$ in the $t-$ channel, with $T=0,1,2$.

\begin{equation}
A_{AB}^{A^\prime B^\prime}=\Gamma_{AA^\prime}A^{(0)}\Gamma_{BB^\prime}+
\Gamma_{AA^\prime}^i A^{(1)}\Gamma_{BB^\prime}^i +
\Gamma_{AA^\prime}^{ij}A^{(2)}\Gamma_{BB^\prime}^{ij}
\label{1.1}
\end{equation}

The constants $\Gamma$
in (\ref{1.1}) depend on the kind of scattering particles
(gauge bosons, fermions, higgs particle),  and
they are all proportional to coupling constant
$\Gamma \propto
 g $, for their explicit
forms see \cite{klf}.

In what follows we shall concentrate ourselves on
 the singlet part of
the amplitude (\ref{1.1}).  $A^{(0)}$ is related to the partial wave
$F_\omega ({ q}^2)$ in the following way ($j=1+\omega$):
\begin{equation}
A^{(0)}(s,q)=\frac{ s}{4 i} \int\limits^{\delta + i\infty}_{\delta - i\infty}
{d\omega}
\left(
\frac{s}{m^2}
\right)^\omega
\frac{e^{-i\pi\omega}-1}{\sin{\pi \omega}}
F_\omega ({ q}^2) \ ,
\quad t=-{ q}^2 \ ,
\label{1.2}
\end{equation}

%$$
%\omega =j-1 \ .
%$$

\begin{equation}
F_\omega ({ q}^2)=\frac{1}{\omega}\frac{g^2}{(2\pi)^3}
\int \frac{d^2 k}{(k^2+m^2)((k-q)^2+m^2)}f_\omega(k,q-k)A_0 (q^2) \ ,
\label{1.3}
\end{equation}

\begin{equation}
A_0 (q^2)=-2(q^2+\frac{5}{4}m^2) \ .
\label{1.4}
\end{equation}
We write here and in the following the scalar products of transverse
momenta in Euclidean notation.
The function $f_\omega(k,q-k)$ satisfies the following integral equation
  (see Fig.1 for notations and graphic form of the equation).

\begin{eqnarray}
& \left[
\omega -\alpha (k^2) -\alpha ((k-q)^2)
\right]
f_\omega (k,q-k) =  & \nonumber \\
&
 \frac{\displaystyle \omega}{\displaystyle A(q^2)}+
\frac{\displaystyle g^2}{\displaystyle  (2\pi)^3}
\int \frac{\displaystyle d^2 k_1}{\displaystyle
(k^2_1+m^2)((k_1-q)^2+m^2)}
K(k,k_1,q)f_\omega (k_1,q-k_1)
 &  \label{1.5}
\end{eqnarray}
with the kernel
\begin{eqnarray}
&
K(k,k_1,q)=A_0 (q^2) &  \\
& +\frac{\displaystyle 2}{\displaystyle (k - k_1)^2 + m^2}
\left[
(k^2+ m^2)((k_1 - q)^2 + m^2)+(k_1^2 + m^2)((k - q)^2 + m^2)
\right] & \label{1.6} \nonumber
\end{eqnarray}

and the Regge trajectory of the masive gluons
\begin{equation}
\alpha(k^2)=j-1=-\frac{g^2(k^2+m^2)}{(2\pi)^3}\int
\frac{d^2k_1}{(k_1^2 + m^2)((k_1 - k)^2 + m^2)} \;\;.
\label{1.7}
\end{equation}

\begin{figure}[htb]
\label{fig-1}
\begin{center}
\epsfig{file=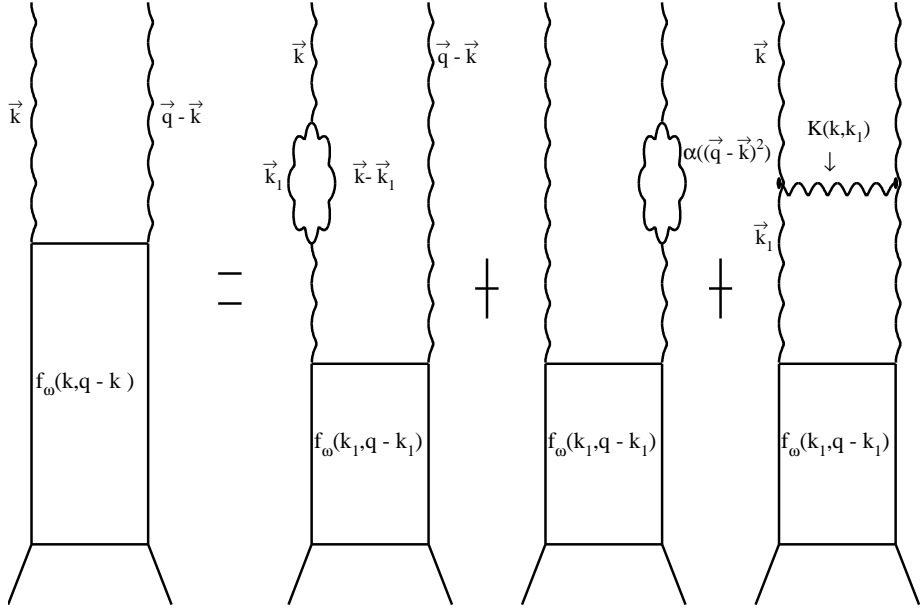,width=14cm}
\caption{The graphic form of the BFKL equation.}
\end{center}
\end{figure}

As a consequence of the integral equation (\ref{1.5}) it is
possible to express the partial wave
$F_\omega ({q}^2)$  through
the solution of Eq. (\ref{1.5}) on the mass--shell

\begin{equation}
f_\omega ({ k},{ k-q})|_{k^2=(q-k)^2=-m^2}=
F_\omega ({ q}^2)+A_0^{-1}(q^2) \ .
\label{1.8}
\end{equation}

\subsection{ 2+1 dimensions}

In Refs. \cite{preprint,marchesini}
it was established that in 2+1 dimensional space--time
the high energy scattering amplitudes derived in LLA are given by 
formulae similar to the ones from the previous section.
 The main difference is that the transverse
space in this case is one--dimensional, so
the  substitution
$$
\frac{d^2 k}{(2\pi )^2}\to\frac{d k}{2 \pi}
$$
should be made.

Therefore
\begin{equation}
F_\omega ({ q}^2)=\frac{1}{\omega}\frac{g^2}{(2\pi)^2}
\int \frac{d k}{(k^2+m^2)((k-q)^2+m^2)}f_\omega(k,q-k)A_0 (q^2) \ ,
\label{2.1}
\end{equation}
where now $g^2$ carries the dimension of mass.

The Regge trajectory for massive vector bosons in the 2+1 dimensions
is given by the  simple
rational expression

\begin{equation}
\alpha(k^2)=-\frac{g^2(k^2+m^2)}{(2\pi)^2}\int
\frac{d k_1}{(k_1^2 + m^2)(k_1 - k)^2 + m^2)}=-\frac{g^2}{2\pi
m}\frac{k^2+m^2}
{k^2+4m^2} \ .
\label{2.2}
\end{equation}

For the function $f_\omega (k, q-k)$ we have here the
one dimensional Bethe--Salpeter type equation

\begin{eqnarray}
& \left[
\omega -\alpha (k^2) -\alpha ((k-q)^2)
\right]
f_\omega (k,q-k) =  & \nonumber \\
&
 \frac{\displaystyle \omega}{\displaystyle A(q^2)}+
\frac{\displaystyle g^2}{\displaystyle  (2\pi)^2}
\int \frac{\displaystyle d k_1}{\displaystyle (k^2_1+m^2)((k_1-q)^2+m^2)}
K(k,k_1,q)f_\omega (k_1,q-k_1) & \label{2.3}
\end{eqnarray}

Our aim is to find the analytic solution of this equation.
It is convenient
to introduce the dimensionless variables
\begin{equation}
\as\,\,=\,\,\frac{g^2}{4\,\pi\,m}\,\,,\,\,\,\,
\epsilon \,=\,\frac{g^2}{2\pi m \omega}\,=\,\frac{2 \as}{\omega}
\;, \;\;\;k \to mk \;,\;\;\;q \to mq
 \ .
\label{2.4}
\end{equation}
Then Eq.(\ref{2.3}) takes the form

\begin{eqnarray}
&
\left[
1+ \epsilon ( \frac{k^2+1}{k^2+4}+ \frac{(k-q)^2+1}{(k-q)^2+4} )
\right]
f_\omega (k,q-k) =  A_0^{-1}+
\epsilon \int \frac{ d k_1}{2\pi }\times & \nonumber \\
&
\left(
\frac{A_0}{ (k^2_1+1)((k_1-q)^2+1)}+
\frac{2}{(k-k_1)^2+1}
\left[
\frac{k^2+1}{k^2_1+1}+
\frac{(k-q)^2+1}{(k_1-q)^2+1}
\right]
\right) f_\omega (k_1,q-k_1) & \label{2.5} \\
&
A_0=-(2q^2+\frac{5}{2}) \label{2.6} \;.
&
\end{eqnarray}
The remaining  equations of the previous section are unchanged.

In order to present the main steps of our method for finding the exact
solution of Eq.(\ref{2.5}) we consider first
 the simpler case with vanishing  momentum transfer $q=0$.

\section{Forward scattering at high energy.}
\setcounter{equation}{0}

In the case with vanishing momentum transfer $q=0$, the \eq{2.5}
takes the simpler form
$$
\frac{(1+ 2\epsilon ) (k^2+\lambda^2)}{k^2+4}
f_\omega (k) -A_0^{-1}=
$$
\begin{equation}
\epsilon
\int \frac{\displaystyle d k_1}{2\pi }
\left(
\frac{A_0}{\displaystyle (k^2_1+1)^2}+
\frac{4}{(k-k_1)^2+1}\frac{k^2+1}{k^2_1+1}
\right) f_\omega (k_1) \ ,
\label{2.1.1}
\end{equation}
where we have introduced  the convenient notation
\begin{equation}
\lambda^2= \frac{4+2\epsilon}{1+2\epsilon} \ .
\label{2.1.2}
\end{equation}
In this case we present the methods of solution both in coordinate and in
momentum representation. In this way different aspects of the problem
will be illuminated.

\subsection{Coordinate space analysis}

We find that it is convenient to work with the function
$\phi_\om(k) = (k^2 +1)^{-1}f_\om(k)$.
Equation  (\ref{2.1.1})  is   a linear
 inhomogeneous integral equation. We try to solve
it in coordinate space by introducing
\beq
\label{c1}
\phi_{\om}(k)\,\,=\,\,\int\,dx \, e^{i k x}\,\phi_{\om}(x)\,\,.
\eeq
The main advantage of the coordinate space is the fact that  the BFKL
kernel  in \eq{2.1.1} looks simple due to the relation
\beq
\label{c2}
 \int \,\frac{d k}{ 2 \pi}\,\,\frac{e^{i kx}
}{k^2\,+\,1}\,\,=\,\,\frac{1}{2}\,e^{- |x|}\,\,.
\eeq
Substituting \eq{c1} and \eq{c2} in \eq{2.1.1} we obtain
\beq
\label{c3}
( 1 \,+\,2 \ep )\,(\, - \frac{d^2}{d^2 x}\,+\,\la^2\,)\,\phi_{\om}(x)
\,\,= \frac{1}{2A_0}(3\,e^{-|x|} +2\delta(x))
\eeq
$$
\,\,+2\,\ep\,(\, - \frac{d^2}{d^2 x}\,+\,4\,)\,e^{-
|x|}\,\phi_{\om}(x) \,  -\, \frac{\ep A_0}{4}\,(
3\,e^{-|x|}\,+\,2\,\de(x)\,)\,\int \,d y \,\phi_{\om}(y)\,e^{- |y|}
$$
where $\de(x)$  is Euler $\de$ - function. We shall analyze Eq.(\ref{c3})
without the inhomogeneous term in order to investigate the leading
eigenvalue $\ep$.

$\phi_\om(x)$ should be bound for the Fourier transform (\ref{c1})
to exist. At large $|x|$ only the left-hand side of (\ref{c3})
is important which leads to the asymptotic solution
$e^{-\la|x|}$.

Clearly the solution depends on $|x|$ only because the kernel
$K(x)$ is an even
function of $x$. We introduce a new function
\beq
\label{c4}
\Phi_{\om}(x)\,\,=\,\,[\,1 \,\,+\,\,2\ep\,(\,1\,\,-\,\,e^{-
|x|}\,)\,]\,\phi_{\om}(x)\,\,.
\eeq
and a new variable $z \,=\, e^{-|x|} $.

For the function $\Phi_\om$ the equation looks as follows
\beq
\label{c5}
- z \frac{d}{d z} \,z\,\frac{d \Phi_{\om}(z)}{d z} \,+\,2 z\,\frac{d
\Phi_{\om}(z)}{d z}\,\de( z - 1)\,+\,4 \,\Phi_{\om}(z)\,\,=
\eeq
$$
\,6\ep\,\frac{\Phi_{\om}(z)}{1 \,+\,2\ep( 1 -
z)}\,\,+ \frac{A_0\,\ep}{4}\,(\,3
z\,+\,2\,\de( z -
1)\,)\,\int^1_0\,d\,z'\,\frac{\Phi_{\om}(z')}{1\,+\,2\ep(\,1\,-\,z'\,)}\,\,.
$$

Comparing the coefficients in front of $\de$-functions we obtain
\beq
\label{c6}
\frac{d \Phi_{\om}(z)}{d z}\,|_{z = 1}\,\,=\,\,\frac{\epsilon A_0}{4}\,\int^1_0\,d
z'\,\frac{\Phi_{\om}(z')}{1 + 2 \ep( 1 - z')}\,\,,
\eeq
which will give the equation for the position of the pole in angular
momentum plane (the intercept of the Pomeron) as will be shown below.
The second condition
\beq
\label{c7}
\Phi_{\om}(z)\,\,\rightarrow\,\,0 \,\,at\,\,\,z\,\,\rightarrow\,\,0
\eeq
follows from the large $|x|$ behaviour of $\phi_\om(x)$ discussed above.

The important observation is that  solution of \eq{c5} obeying
(\ref{c7})  can be
found in
the form
\beq
\label{c8}
\Phi_{\om}(z)\,\,=\,\,C z [1 + 2\ep( 1 -
z)]\,\,+\,\,\Phi^{hg}_{\om}(z)\,\,,
\eeq
where $\Phi^{hg}_{\om}(z)$ is the solution of the homogeneous equation
\beq
\label{c9}
- z \frac{d}{d z} \,z\,\frac{d \Phi^{hg}_{\om}(z)}{d z} \,\,+\,\,4
\,\Phi^{hg}_{\om}(z)\,\,=
\,6\ep\,\frac{\Phi^{hg}_{\om}(z)}{1 \,+\,2\ep( 1 -
z)}\,\,,
\eeq
and  C  is a constant in $z$, which however depends on the
function $\Phi_\om$,
\beq
\label{c10}
C\,\,=-\,\,\frac{5 \,\ep}{8 + \frac{5}{2} \,\ep}\,\int^1_0 \,d
z'\,\frac{\Phi^{hg}_{\om}}{1 \,+ \,2\,\ep ( 1 - z')} \,\,.
\eeq

The solution of the homogeneous equation  (\ref{c9})  can be easily
found. We obtain,
\beq
\label{c11}
\Phi^{hg}_{\om}(z)\,\,=\,\,N\,z^{\la}\,\,{}_2F_1(2 + \la, -2 + \la, 1 +
2\la,\frac{2 \ep z}{1 + 2\ep})\,\,=
\eeq
$$
\,\,N\,\frac{( 1 +
2 \ep (1 - z)) \,z^{\la}}{1 + 2 \ep }\,\,{}_2F_1(3 +  \la, -1 +
\la, 1 + 2\la,\frac{2 \ep z}{1 + 2\ep})\,\,.
$$
Here   ${}_2F_1$ is the Gauss hypergeometric  function and the
constant
$N$ can be defined from the normalization.

To find the value of $\ep$ which corresponds to the bound state we have to
solve \eq{c6} which using well known properties of the hypergeometric
function can be  reduced to the form
\beq
\label{c12}
 \frac{5 - 2 \la^2}{4 - \la^2}{}_2F_1(3 +  \la, -1 +\la, 1 + 2\la,
\frac{4 - \la^2}{3}) \;+
\eeq
$$
 \frac{( \la^2 - 1)( \la - 1)}{\la^2 - 4}
{}_2F_1(3 +  \la, \la, 1 + 2\la,\frac{4 - \la^2}{3})\,\,=
$$
$$
\frac{2}{ 1 + \frac{16}{5} \,\frac{1 - \la^2}{\la^2 - 4}}\,\frac{1}{2 +
\la}\,(\, {}_2F_1(2 +  \la, -1 +
\la, 1 + 2\la,\frac{4 - \la^2}{3})\,
$$
$$
+\,\frac{1}{1 + \la} \,
{}_2F_1(1 +  \la, -1 + \la, 1 + 2\la,\frac{4 - \la^2}{3})\,)
$$

We solved this equation numerically and obtained
 the value $\ep = \ep_0\,=4.5934$ which leads to the rightmost singularity
at $\om\,=\,\om_0\,\,=\,\,\frac{2 \as}{\ep_0}\,\,=\,\,0.436
\,\as$ in accorance with ref. \cite{preprint}.

The way we have solved the BFKL equation is reminiscent of the
standard procedure of
calculating bound states. The rightmost
singularity in $\om$, a pole, corresponds to the ground state.
In the following subsection we solve \eq{2.1.1} in momentum representation.

\subsection{ Momentum space analysis.}

We have to solve the linear inhomogeneous integral equation
(\ref{2.1.1}). It is possible to construct the solution directly by
iterations. We rely on the relation
\beq
\label{example}
\int\limits^{\infty}_{-\infty} \frac{dk'}{2\pi}\,\frac{1}{(k-k')^2 +
1}\frac{1}{k'^2
+\la^2} = \frac{\la +1}{2\la [k^2 + (\la +1)^2]}
\eeq
This means that the action of the kernel on $(k^2 + \la^2)^{-1}$
can be expressed by the shift of the pole position $\la \to \la + 1$.

Let us formally consider the right--hand side of Eq. (\ref{2.1.1})
as a perturbation. We will consider first the solution of
this equation in the interval
 $\epsilon \in [-1/2,\infty]$, where $\lambda^2>0$.
We obtain the solution for $\la > 0$ first and continue then analytically
to the complete complex plane in $\epsilon$ or $\om$.

If we omit the right--hand side of Eq.(\ref{2.1.1}) (the zeroth
iteration),
the solution is

\begin{equation}
f_\omega^{(0)} (k)=\frac{A_0^{-1}(k^2+4)}{(1+2\epsilon )(k^2 + \lambda^2)} \ .
\label{2.1.3}
\end{equation}

$f_\omega^{(0)} (k)$  can be represented as the sum of the constant term plus
the pole term $\sim \frac{1}{k^2+\lambda^2}$.
In order to find the contribution arising from the next iteration
($f_\omega=f_\omega^{(0)}+
f_\omega^{(1)}+ \dots$) let us substitute (\ref{2.1.3}) into the
right--hand side
of (\ref{2.1.1}). We use now \eq{example} and obtain
\begin{eqnarray}
&& \int\limits^{\infty}_{-\infty} K(k,k_1,q=0) {1 \over k_1^2+\lambda^2}
{dk_1 \over 2\pi} =  \nonumber \\
&& {A_0(\lambda +2) \over 4\lambda  (\lambda +1)^2} +
{2(k^2+1) \over \lambda (k^2+4)}\left[
{1 \over \lambda +1}+ {\lambda +3 \over k^2 +(\lambda +1)^2}
\right]
\end{eqnarray}
We write
the resulting first iteration
$f_\omega^{(1)}$  as a sum of pole terms in $k^2$.
In this expansion there are three  terms: the constant term, the pole term
$\sim \frac{1}{k^2+\lambda^2}$ and as a new term, not encountered in the
zeroth iteration
 $f_\omega^{(0)}$, the pole term
$\sim \frac{1}{k^2+(\lambda +1)^2}$.
The same procedure can be applied to the subsequent iterations.
It is easy to see that the expansion for the n-th iteration will be given
by the
sum of the constant term plus the pole terms
$\frac{A_k}{k^2+\lambda_k^2}, \lambda_k=
\lambda+k-1, k=1,\dots, n+1$.
Therefore it is natural to look for the solution of Eq. (\ref{2.1.1}) in
the form \cite{preprint}

\beq
\label{2.1.4}
f_\omega (k)= f_0 + \sum_{n=1}^\infty
\frac{A_n}{k^2+\lambda_n^2} \ , \ \lambda_n=\lambda +n-1 \ .
%\label{2.1.4}
\eeq

Let us substitute this ansatz in the equation (\ref{2.1.1}).
Comparing coefficients of the pole terms
on both sides we find the condition

\begin{equation}
\frac{A_n}{A_{n-1}}=\frac{2\varepsilon}{1+ 2\varepsilon}
\frac{(\lambda +n)
(\lambda + n+1)}{(n-1)
(2\lambda+n+1)} \ .
\label{2.1.5}
\end{equation}

This recurrence relation has the following solution

\begin{equation}
A_n=A_1\left(
\frac{2\varepsilon}{1+2\varepsilon}
\right)^{n-1}\frac{(\lambda_1+2)_{n-1}(\lambda_1+3)_{n-1}}{(n-1)!
(2\lambda_1+1)_{n-1}} \ ,
\label{2.1.6}
\end{equation}
where $(a)_n=a(a+1)(a+2)\dots(a+n-1)$. In this way we arrive at
generalized hypergeometric functions
${}_{p+1}F_p(^{\al_1 \dots \alpha_{p+1}}_{\be_1 \dots
\beta_p}|y)$ \cite{kirpich}.
In particular, the ansatz (\ref{2.1.4}) leads to

\begin{equation}
 f_\omega (k) =f_0 +
 \frac{A_1}{k^2+ \lambda^2}
{}_4F_3(^{\lambda_3,\lambda_4,\lambda+ik,\lambda-ik}_{
2\lambda+1,\lambda_2+ik,\lambda_2-ik}|y) \ .
\label{2.1.7}
\end{equation}
with  $\lambda_n=\lambda+n-1$, $\lambda=\lambda_1$ and
$y=\frac{\displaystyle 2\varepsilon}{\displaystyle 1+2\varepsilon}$.

There are still two coefficients $f_0$ and $A_1$ undetermined in our
solution (\ref{2.1.7}). The information contained in equation
(\ref{2.1.1})
which has not been used yet can be expressed in terms of two conditions.
The first condition appears as a result of the comparison of residua of
the  pole term $\sim \frac{1}{k^2+\lambda^2}$
( the pole at $k^2 \to \lambda^2$
has to be considered separately from other pole terms $\sim
\frac{1}{k^2 + \lambda_n^2}$, $n \neq 1$).
The second condition appears as a result of the comparison of the
constant terms appearing in
the expansion,
or in other words, considering the left and the right--hand sides of
Eq. (\ref{2.1.1}) at $k\to \infty$.

Therefore the coeficients $f_0, A_1$
are the solution of the following in\-ho\-mo\-ge\-neous
system of  linear equations

\begin{eqnarray}
&-\frac{1}{A_0 \epsilon} = f_0 \cdot a_{11}+A_1 \cdot a_{12}&
\nonumber \\
&-\frac{1}{A_0 \epsilon} = f_0 \cdot a_{21}+A_1 \cdot a_{22} & \ ,
\label{2.1.8}
\end{eqnarray}
with
\[
a_{11}= \frac{A_0}{4}+\frac{4(\lambda^2-1)}{(\lambda^2-4)} \nonumber
\label{2.1.9}
\]

\begin{eqnarray}
& a_{12}=-\frac{6}{(\lambda^2-4)^2}+
\frac{A_0\lambda_3}{4\lambda\lambda_2^2}
{}_4F_3(^{\lambda_4,\lambda_4,\lambda,\lambda_2}
_{2\lambda +1,\lambda_3,\lambda_3}
|y) & \nonumber \\
&  + \frac{4}{(\lambda^2-4)}{}_3F_2(^{\lambda_4,\lambda_2,\lambda-1}
_{2\lambda +1,\lambda}
|y)+\frac{2\lambda_2}{\lambda(2\lambda+1)}
{}_4F_3(^{\lambda_4,\lambda_3,\lambda-1,1}
_{2\lambda +2,\lambda_2,2}
|y) &
\label{2.1.10}
\end{eqnarray}

\[
a_{21}=\frac{A_0}{4}+\frac{2(2\lambda^2-5)}{(\lambda^2-4)}
\label{2.1.11}
\]

\[
a_{22}=
\frac{A_0\lambda_3}{4\lambda\lambda_2^2}
{}_4F_3(^{\lambda_4,\lambda_4,\lambda,\lambda_2}
_{2\lambda +1,\lambda_3,\lambda_3}
|y)
+ \frac{2}{\lambda\lambda_2}{}_2F_1(^{\lambda_4,\lambda}
_{2\lambda +1}
|y) \nonumber
\label{2.1.12}
\]

Using well known  relations among the hypergeometric functions
\cite{kirpich}
it is possible to express all higher hypergeometric functions
through the two basic ${}_2F_1$ functions
\[
f_a={}_2F_1(^{\lambda_2,\lambda}
_{2\lambda +1}
|y)
\label{2.1.13}
\]

\begin{equation}
f_b={}_2F_1(^{\lambda ,\lambda}
_{2\lambda +1}
|y)
\label{2.1.14}
\end{equation}
We quote here only one of these relations
\begin{equation}
{}_4F_3(^{\lambda_4,\lambda_4,\lambda,\lambda_2}
_{2\lambda +1,\lambda_3,\lambda_3}
|y)
=f_a\frac{(7\lambda^2-4)}{(\lambda-1)\lambda_2\lambda_3^2}+
f_b\frac{9\lambda^3}{(\lambda-1)^2\lambda_2^2\lambda_3^2}
\label{2.1.15} \;.
\end{equation}

The solution of the system (\ref{2.1.1}) expressed in terms of the functions
$f_a$ and $f_b$ has the form

\begin{eqnarray}
& A_1=-\frac{\lambda_2^2\lambda_3^2(\lambda-2)}{A_0\epsilon }
\left[\frac{A_0}{4}(f_a\frac{(13\lambda^2-16)}{\lambda}+18f_b)+
\right. & \nonumber \\
& \left. f_a\frac{(34\lambda^2-64)(\lambda^2-1)}{\lambda (\lambda-2)\lambda_3}+
f_b\frac{48\lambda^2-84}{(\lambda-2)\lambda_3} \right]^{-1} &
\label{2.1.16} \nonumber
\end{eqnarray}

\begin{equation}
f_0=\frac{3A_1}{(\lambda-1)(\lambda-2)\lambda_3^2\lambda_2^3}
(f_a \lambda(1+2\lambda^2)
+f_b\frac{3(\lambda^4+2)}{(\lambda-1)\lambda_2})
\label{2.1.17}
\end{equation}

These formulae together with Eq. (\ref{2.1.7}) represent
 the solution of the integral equation
(\ref{2.1.1}).

It should be noted that in Ref. \cite{preprint} instead of the first equation
of the  system (\ref{2.1.8}) (resulting from the comparision of the
residua of the pole terms $\frac{1}{k^2+\lambda^2}$ appearing on both sides
of eq.(\ref{2.1.1}))
another boundary condition was used, the absence of the normal
thresholds,
\begin{equation}
f_\omega(k^2\to -4)=2f_\omega(k^2\to -1) \ .
\label{2.1.18}
\end{equation}
This condition can be derived from the equation (\ref{2.1.1})
if one requires that $f_\omega (k) $ is a
regular function in the neigbourhood of  $k^2 = -4$.
In terms of our ansatz this condition reads
\begin{equation}
f_0=A_1\left[
\frac{1}{\lambda^2-4} {}_3F_2(^{\lambda_3,\lambda_3,\lambda-2}
_{2\lambda+1,\lambda-1} |y)-\frac{2}{\lambda^2-1}
{}_3F_2(^{\lambda_4,\lambda_2,\lambda-1}
_{2\lambda+1,\lambda} |y) \right]
\label{2.1.19}
\end{equation}

The iterative solution of the equation (\ref{2.1.1}),
$f_\omega (k)$, as described above,
is a function which is by construction regular in the points
$k^2 = -n^2$. Therefore, the condition (\ref{2.1.19})
should not give an additional restriction on the function (\ref{2.1.7})
as compared with the conditions given by the system (\ref{2.1.8}).
Indeed, expressing the hypergeometric functions in (\ref{2.1.19})
in terms of the functions $f_a$ and $f_b$ it can be checked directly
that the difference
of the two equations in  (\ref{2.1.8}) and the condition
(\ref{2.1.19})
are equivalent.

\subsection{Regge singularities of the forward partial wave}

We discuss now the implications of the obtained solution
 $f_\omega (k)$ for the partial wave of the scattering amplitude $F_\om$.
 The partial wave $F_\om$ can be calculated either by (\ref{2.1})
or by the mass-shell relation (\ref{1.8}).
We have checked that the both methods lead to  the same result
\begin{equation}
F_\omega (q=0)=\frac{-1}{A_0+f} \ ,
\label{2.1.20}
\end{equation}
where
\begin{equation}
f=\frac{4}{\lambda^2-4}
\left(\frac{(34\lambda^2-64)(\lambda^2-1)}{\lambda}f_a+
f_b(48\lambda^2-84)\right)\left(\frac{(13\lambda^2-16)}{\lambda}f_a+
18f_b\right)^{-1}
\label{2.1.21}
\end{equation}

Let us discuss the singularities of $F_\omega$ considered as a function of
the complex variable $\omega$. The hypergeometric
functions are defined in terms of the hypergeometric
series which are convergent inside
the circle of unit radius in the variable $y=2\epsilon/(1+2\epsilon)$.
The continued hypergeometric functions are analytic in the complex plane
of their argument $y$, with a cut from $y=1$ to $y=\infty$.
 In the $\epsilon$ plane this corresponds to the cut appearing on the
interval
$\epsilon \in [-\infty, -1/2]$.

As a function of their parameters $\al_1,\dots \al_{p+1}, \be_1, \dots
\be_p$ the hypergeometric functions have only simple poles if one of the
lower parameters $\be_1, \dots \be_p$ approaches a non-positive integer
value $n$. We see from (\ref{2.1.13}), (\ref{2.1.14}) that both
$f_a$ and $f_b$ have poles of this origin at
 $\lambda=\sqrt{\frac{4+2\epsilon}{1+2\epsilon}}\to -\frac{1+n}{2}$.
Therefore
these poles lie on the second (unphysical) sheet of the square root.

Further singularities of $f_\om$ and, consequently, of $F_\om$ appear at
points, where the determinant of the system of linear equations
(\ref{2.1.8}) vanishes, i.e. at the zeros of the denominator in
(\ref{2.1.20})
\begin{equation}
A_0+f=0 \
.
\label{2.1.22}
\end{equation}
This results in  poles in $\om$.

Analyzing the condition (\ref{2.1.22})
numerically outside the interval $\epsilon \in [-\infty, -1/2]$, where
the cut is located we have checked that there is only one Regge pole in
the vicinity of the real axis located in
\begin{equation}
\epsilon=\epsilon_0=4.5934 \ ,
\ \omega=\omega_0 =\frac{g^2}{2\pi m \epsilon_0} \ .
\label{2.1.23}
\end{equation}
The result coincides of course with the one obtained in the coordinate
representation.
Therefore we can conclude that at $q=0$ the partial wave $F_\omega$
has the following singularities on the physical sheet of complex
$\omega$ plane. There is a finite cut
on the negative part of the real axies covering the interval
$\omega \in [\omega_2,\omega_1]$, with
$\omega_2=\frac{-g^2}{\pi m}$, $\omega_1=0$. And there is a single pole
in the positive part of the real axis at $\omega=\omega_0$,
(\ref{2.1.23}).

Let us discuss the nature of the singularities at the branch points.
Near the
 right end-point of the cut, $\omega=\omega_1=0$ we have
$\epsilon \to +\infty$, $\lambda \to 1$, $y\to 1$ and

\[
f_a= -2(1+\log{\frac{2(\lambda -1)}{3}})+ O((\lambda-1)\log(\lambda -1))
\label{2.1.24} \;,
\]

\begin{equation}
f_b= 2+ O((\lambda-1)\log(\lambda -1))
\label{2.1.25} \;.
\end{equation}

Therefore the partial wave behaves like

\begin{equation}
F_\omega\to -A_0^{-1}+16A_0^{-2}/\log(\lambda -1)
\label{2.1.26}
\end{equation}

Near the left end-point of the cut, $\omega=\omega_2$ we have
$\epsilon \to -1/2$, $\lambda \to +\infty$,
$y\sim -\lambda^2/3 \to - \infty$ and

\begin{equation}
f_a= \frac{e^{\lambda (\log{12}-2\log{\lambda})}}
{ \sqrt{\pi \lambda}}(1+  O(1/\lambda))
\label{2.1.27} \nonumber
\end{equation}

\begin{equation}
f_b= \frac{e^{\lambda (\log{12}-2\log{\lambda})}\lambda\log{\lambda}}
{ \sqrt{\pi \lambda}}(1+  O(1/\lambda))
\label{2.1.28}
\end{equation}

Therefore the partial wave behaves like

\begin{equation}
F_ \omega\to -\frac{1}{A_0+24/3}-\frac{1}{72 \cdot
(A_0+24/3)^2\log{\lambda }}
\label{2.1.29}
\end{equation}

Note that the solution of the corresponding homogeneous
equation can be obtained from the solution of inhomogeneous
equation $F_\omega$ which we have just found.
The spectrum consists of one discret
level $\omega=\omega_0$ and the continuous part $\omega \in
[\omega_2,\omega_1]$.
The corresponding eigenfunctions can be found as follows:
the residue of $F_\omega$ of the pole at $\omega = \omega_0 $
gives (up to the normalization constant) the wave function of the
discret level and by calculating the discontinuity of $F_\omega$ on the
cut
it is possible to find the eigenfunctions belonging to the
continuous spectrum.

We would like to add a comment on how the results depend on the number of
colours $N$. In the case of arbitrary $N$ we have to substitute in the above
equations \cite{klf}
\beqar{N}
&&g^2 \to g^2\,\frac{N}{2}\;,\;\;\;\; A_0 \to -2(q^2
+ \frac{N^2 + 1}{N^2}\,m^2) \nonumber \\
&& \epsilon \to \epsilon\,\frac{N}{2}\;\;.
\eeqar
As in the case $N=2$ there is a leading Regge pole at arbitrary $N$ located
at $\om_0^{(N)}$,
\beq
\om_0^{(N)} = \frac{g^2}{2\pi m}\cdot \frac{N}{2}\cdot
\frac{1}{\epsilon_0^{(N)}} \;\;.
\label{omegaN}
\eeq
$\epsilon_0^{(N)}$ is calculated in analogy to $\epsilon_0$ above.
We find that $\epsilon_0^{(N)}$ decreases slowly with $N$ approaching
a limit $\epsilon_0^{(\infty)}$:
$\epsilon_0^{(2)} = 4,5934$,
$
\epsilon_0^{(3)}=3.8000$, $\epsilon_0^{(4)}=3.5693$,
$\epsilon_0^{(\infty)}= 3.3025$.

\section{Non-forward scattering}
\setcounter{equation}{0}

\subsection{Solution of the equation}

The main steps which have been made in the
 section 3.2 to derive the solution of the forward equation
can be generalized to find the solution of the non-forward equation
(\ref{2.5}).
The expression appearing on the left-hand side of Eq.(\ref{2.5})
in the square brackets $[\dots]$ can be rewritten in the form

\begin{equation}
[\dots ]=\frac{(1+2\epsilon)[x^2+(\lambda^+)^2][x^2+(\lambda^-)^2]}
{[(x - q/2)^2 + 4][(x+ q/2)^2 +4]}
\label{2.2.1}
\end{equation}
where
\[
x=k-q/2 \ ,
\label{2.2.2}
\]

\begin{equation}
(\lambda^{\pm})^2=\frac{4+5\epsilon}{1+2\epsilon}-\frac{q^2}{4}
\pm \sqrt{\frac{9\epsilon^2-q^2(4+5\epsilon)(1+2\epsilon)}
{(1+2\epsilon)^2}}  \ .
\label{2.2.3}
\end{equation}

Now, in  analogy with the iterative way of finding the
solution for $q=0$ , we see,
that the zeroth iteration can be expanded into a sum of the constant term,
and two pole terms: $\sim \frac{1}{x^2+(\lambda^+)^2}$ and
$\sim \frac{1}{x^2+(\lambda^-)^2}$.
It should be noted that the zeroth iteration for  $f_\omega (k,q-k)$
 depends on the specific combination of the momenta
$k$ and $q-k$, i.e. it is a function of the variable $x^2=
(k-q/2)^2$. The notation $x$ should not be confused with the position.
Calculating the next iterations it can be seen
that this feature remains true and the solution can be represented
in the following form \cite{preprint}

\begin{equation}
f_\omega (x^2)=f_0 + \sum^\infty _{n=1} \frac{A_n^+}{x^2+(\lambda^+_n)^2}
+ \sum^\infty _{n=1} \frac{A_n^-}{x^2+(\lambda^-_n)^2} \ , \
\lambda^{\pm}_n=\lambda^\pm + n -1 \ .
\label{2.2.4}
\end{equation}

Substituting this ansatz into the equation (\ref{2.5}) we  find two
 recurrence relations similar to the one for the case $q=0$.
 Their solutions can be written in the form
\begin{eqnarray}
&
\frac{A_n^\pm}{A^\pm_1}=\left(
\frac{2\epsilon}{1+ 2\epsilon}
\right)^{n-1}\frac{1}{(n-1)!}\times
 &\nonumber \\
&
\frac{(\lambda_2^\pm)_{n-1} (\lambda_4^\pm+\frac{iq}{2})_{n-1}
(\lambda_4^\pm-\frac{iq}{2})_{n-1} (\lambda^\pm-d^+_+)_{n-1}
(\lambda^\pm-d^+_-)_{n-1} (\lambda^\pm-d^-_+)_{n-1}
(\lambda^\pm-d^-_-)_{n-1}}
{(\lambda^\pm)_{n-1} (2\lambda^\pm+1)_{n-1}
(\lambda^\pm+\lambda^{\mp}+1)_{n-1} (\lambda^\pm-\lambda^{\mp}+1)_{n-1}
(\lambda_2^\pm+\frac{iq}{2})_{n-1}
(\lambda_2^\pm-\frac{iq}{2})_{n-1} }&
\label{2.2.5}
\end{eqnarray}
with
\begin{equation}
d_{a}^{b}=-\frac{1}{2} + a
\frac{1}{2}\sqrt{5-q^2 + 2b \sqrt{4-3q^2}}, \;\;\;\;a, b = \pm \ .
\label{2.2.5b}
\end{equation}

It is possible to rewrite our ansatz Eq.(\ref{2.2.4}) in terms of the
generalized
hypergeometric functions

\begin{eqnarray}
&&
f_\omega (x^2)= f_0 +
 \nonumber \\
&&
A_1^+
\frac{1}{x^2+ (\lambda^+)^2}\;
{}_9F_8(
^{\lambda_2^+,\lambda_4^+ +\frac{iq}{2},
\lambda_4^+ -\frac{iq}{2},
\lambda^+ -d^+_+,\lambda^+ -d^+_-,\lambda^+ -d^-_+,\lambda^+ -d^-_-,
\lambda^+ +ix,\lambda^+ -ix}
_{\lambda^+, 2\lambda^+ +1, \lambda^+ + \lambda^- +1,
\lambda^+ - \lambda^- +1,\lambda_2^+ +\frac{iq}{2},
\lambda_2^+ -\frac{iq}{2},\lambda^+_2 +ix,\lambda^+_2 -ix}|y)
+  \nonumber \\
&&A_1^- \frac{1}{x^2+ (\lambda^-)^2}
{}_9F_8(\lambda^+ \leftrightarrow \lambda^- |y) \ .
\label{2.2.6}
\end{eqnarray}

The conditions which determine the coefficients $f_0$, $A_1^+$, $A_1^-$
are also analogous to the ones used in the case of $q=0$.

The condition obtained by taking the limit $x \to \infty$, or $k \to
\infty$,
has the form

\begin{eqnarray}
&
- \frac{1}{A_0\epsilon}=
f_0\left[\frac{A_0}{q^2+4} +\frac{2\epsilon -1}{\epsilon} \right] +
& \nonumber \\
&
A_1^+ \left[
\frac{A_0\lambda_3^+}{\lambda^+ (q^2+4)((\lambda_2^+)^2+\frac{q^2}{4})}
{}_7F_6
(
^{\lambda_4^+,\lambda_4^+ +\frac{iq}{2},
\lambda_4^+ -\frac{iq}{2},
\lambda^+ -d^+_+,\lambda^+ -d^+_-,\lambda^+ -d^-_+,\lambda^+ -d^-_-}
_{\lambda^+_3, 2\lambda^+ +1, \lambda^+ + \lambda^- +1,
\lambda^+ - \lambda^- +1,\lambda_3^+ +\frac{iq}{2},
\lambda_3^+ -\frac{iq}{2}}
|y) + \right. & \nonumber \\
&
\left.
\frac{2\lambda_2^+}{\lambda^+ ((\lambda_2^+)^2+\frac{q^2}{4})}
{}_7F_6
(
^{\lambda_3^+,\lambda_4^+ +\frac{iq}{2},
\lambda_4^+ -\frac{iq}{2},
\lambda^+ -d^+_+,\lambda^+ -d^+_-,\lambda^+ -d^-_+,\lambda^+ -d^-_-}
_{\lambda^+_2, 2\lambda^+ +1, \lambda^+ + \lambda^- +1,
\lambda^+ - \lambda^- +1,\lambda_3^+ +\frac{iq}{2},
\lambda_3^+ -\frac{iq}{2}}
|y) \right] & \nonumber \\
&
+ A_1^-[ \lambda^+ \leftrightarrow \lambda^-] \ .
&
\label{2.2.7}
\end{eqnarray}

We have found that  it is convenient to use as the last two
 conditions  the  absence
of normal thresholds (see discussion at the end of of section 3.2)

\begin{equation}
f_\omega ((\frac{q}{2}\pm 2i)^2)=2f_\omega ((\frac{q}{2}\pm i)^2) \ .
\label{2.2.8}
\end{equation}

The condition corresponding to the lower signs is

\begin{eqnarray}
&
f_0=
A_1^+
\left[
\frac{1}{((\lambda^+)^2-(2+\frac{iq}{2})^2)}
{}_8F_7
(
^{\lambda_2^+,\lambda_3^+ +\frac{iq}{2},
\lambda_4^+ -\frac{iq}{2},\lambda^+ -2  -\frac{iq}{2},
\lambda^+ -d^+_+,\lambda^+ -d^+_-,\lambda^+ -d^-_+,\lambda^+ -d^-_-}
_{\lambda^+, 2\lambda^+ +1, \lambda^+ + \lambda^- +1,
\lambda^+ - \lambda^- +1,\lambda_2^+ +\frac{iq}{2},
\lambda_2^+ -\frac{iq}{2},\lambda^+ -1 -\frac{iq}{2}}
|y) - \right. & \nonumber \\
&
\left.
\frac{2}{((\lambda^+)^2-(1+\frac{iq}{2})^2)}
{}_8F_7
(
^{\lambda_2^+,\lambda_4^+ +\frac{iq}{2},
\lambda_4^+ -\frac{iq}{2},\lambda^+ -1 -\frac{iq}{2},
\lambda^+ -d^+_+,\lambda^+ -d^+_-,\lambda^+ -d^-_+,\lambda^+ -d^-_-}
_{\lambda^+, 2\lambda^+ +1, \lambda^+ + \lambda^- +1,
\lambda^+ - \lambda^- +1,\lambda_3^+ +\frac{iq}{2},
\lambda_2^+ -\frac{iq}{2},\lambda^+ -\frac{iq}{2}}
|y) \right] & \nonumber \\
&
+ A_1^-[ \lambda^+ \leftrightarrow \lambda^-] \ .
&
\label{2.2.9}
\end{eqnarray}
The other equation is obtained from the above one by the
substitution $q \leftrightarrow -q$.

The difference of these two condition can be written in the
limit $q\to 0$ as

\begin{equation}
A_1^+
\left\{
\frac{1}{-iq} + O(1)
\right\}
+
A_1^-\left\{
-iCq + O(q^2)
\right\}=0
\label{2.2.10}
\end{equation}

Therefore

\begin{equation}
A_1^+=A_1^- [Cq^2 + O(q^3)] \ ,
\label{2.2.11}
\end{equation}
where C is some constant.

We see that at $q\to 0$ the series of poles $\sim \frac{A_n^+}{x^2+
(\lambda_n^+)^2}$  decouples
 from the solution in accordance with our previous
considerations  for $q=0$.

In this way we have solved the equation (\ref{2.3}) for arbitrary momentum
transfer $q$. The solution $f_\om$ is given by (\ref{2.2.6}) with the
coefficients $f_0$, $A_1^+$ and $A_1^-$ determined from linear system of
equations (\ref{2.2.7}), (\ref{2.2.8}-\ref{2.2.9}).

\subsection{Properties of the partial wave}

We investigate the partial wave $F_\om(q^2)$ obtained from the solution by
 Eq. (\ref{2.1})

\begin{eqnarray}
&
F_\omega(q^2)=\frac{\epsilon A_0}{q^2+4}
\left[ f_0 +
\frac{A_1^+\lambda_3^+}
{\lambda^+ [(\lambda_2^+)^2+\frac{q^2}{4}]}
{}_7F_6(
^
{
\lambda^+_4,\lambda^+_4+\frac{iq}{2},
\lambda^+_4-\frac{iq}{2},
\lambda^+ -d^+_+,\lambda^+ -d^+_-,\lambda^+ -d^-_+,\lambda^+ -d^-_-
}
_{
\lambda^+_3,\lambda^+_3+\frac{iq}{2},
\lambda^+_3-\frac{iq}{2},
2\lambda^+ +1,\lambda^+ +\lambda^- +1,\lambda^+ -\lambda^- +1
}
|y)
\right.
&
 \nonumber \\
&
\left.
+
\frac{A_1^-\lambda_3^-}
{\lambda^-[(\lambda_2^-)^2+\frac{q^2}{4}]}
{}_7F_6(
^
{
\lambda^-_4,\lambda^-_4+\frac{iq}{2},
\lambda^-_4-\frac{iq}{2},
\lambda^- -d^+_+,\lambda^- -d^+_-,\lambda^- -d^-_+,\lambda^- -d^-_-
}
_{
\lambda^-_3,\lambda^-_3+\frac{iq}{2},
\lambda^-_3-\frac{iq}{2},
2\lambda^- +1,\lambda^+ +\lambda^- +1,\lambda^- -\lambda^+ +1
}
|y)
\right] \;\;.
 & \label{solution-partial}
\end{eqnarray}

First of all it should be noted that all  equations above
are written under the assumption that we choose the convention
for the square root expression for $\lambda^\pm$ with
 the real parts of $\lambda^\pm$ being positive for small $q$ and real
$\epsilon>-\frac{1}{2}$.

If $q^2\geq \frac{9}{10}$,
$\lambda^\pm$ are complex conjugate numbers
if $\epsilon$ belongs to the interval $[\epsilon_1,\infty]$, where
$\epsilon_1 =\frac{13q^2-3q\sqrt{q^2+16}}{2(9-10q^2)}$.
Since $-1/2<\lambda_1<0$, for any positive $\omega$
and therefore for any positive $\epsilon$,
$\lambda^\pm$ are complex conjugate.
Let us choose by the definition as  $\lambda^+$ the expression which has
negative imaginary part (for $q>0$).

In the following we study the Regge singularities and the behaviour at
$-q^2 =t \to - \infty$ and at positive $t$ up to the first threshold
$t=4$.

Since the solution behaves smooth at $q \to 0$ we conclude that at small
$q$ the structure of the Regge singularities is similar to what we have
found for $q=0$. The position of the leading Regge pole depends on $t = -q^2$.
The result of the numerical calculations is plotted in Fig.~2
for values of $t$ from $-4$ up to the vicinity of the first threshold
at $t=4$.
The trajectory is almost linear in the vicinity of $t=0$ with the
approximate slope 0.34$\frac{\as}{m^2}$ as shown in Fig.~3 .

\begin{figure}
\label{traj}
\begin{center}
\epsfig{file=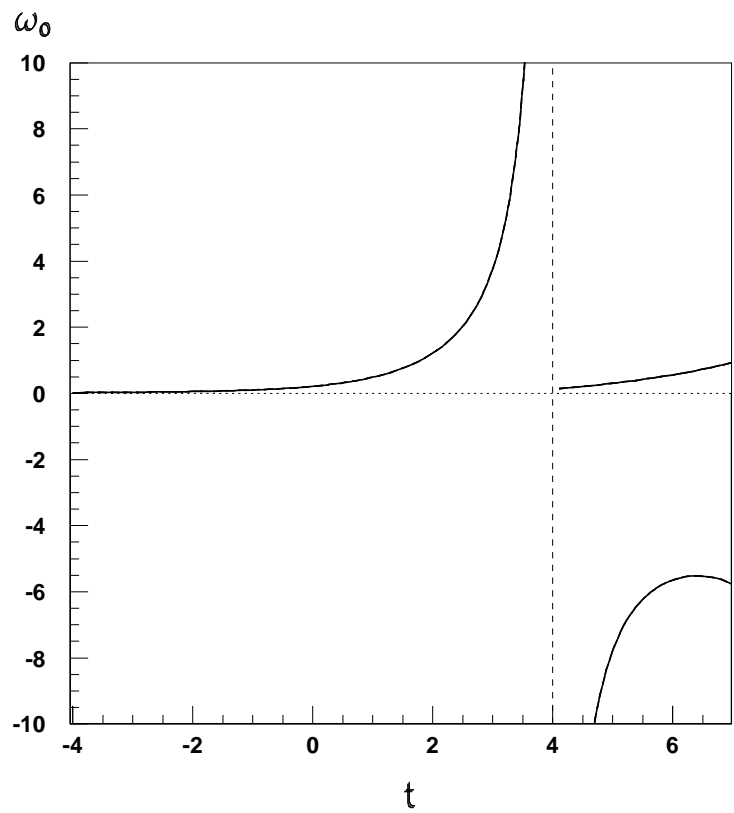,width=12cm}
\caption{The trajectory of the pomeron pole in units of
$\frac{g^2}{2\pi m}$. The momentum transfer is given in units of $m^2$.}
\end{center}
\end{figure}

\begin{figure}
\label{traj0}
\begin{center}
\epsfig{file=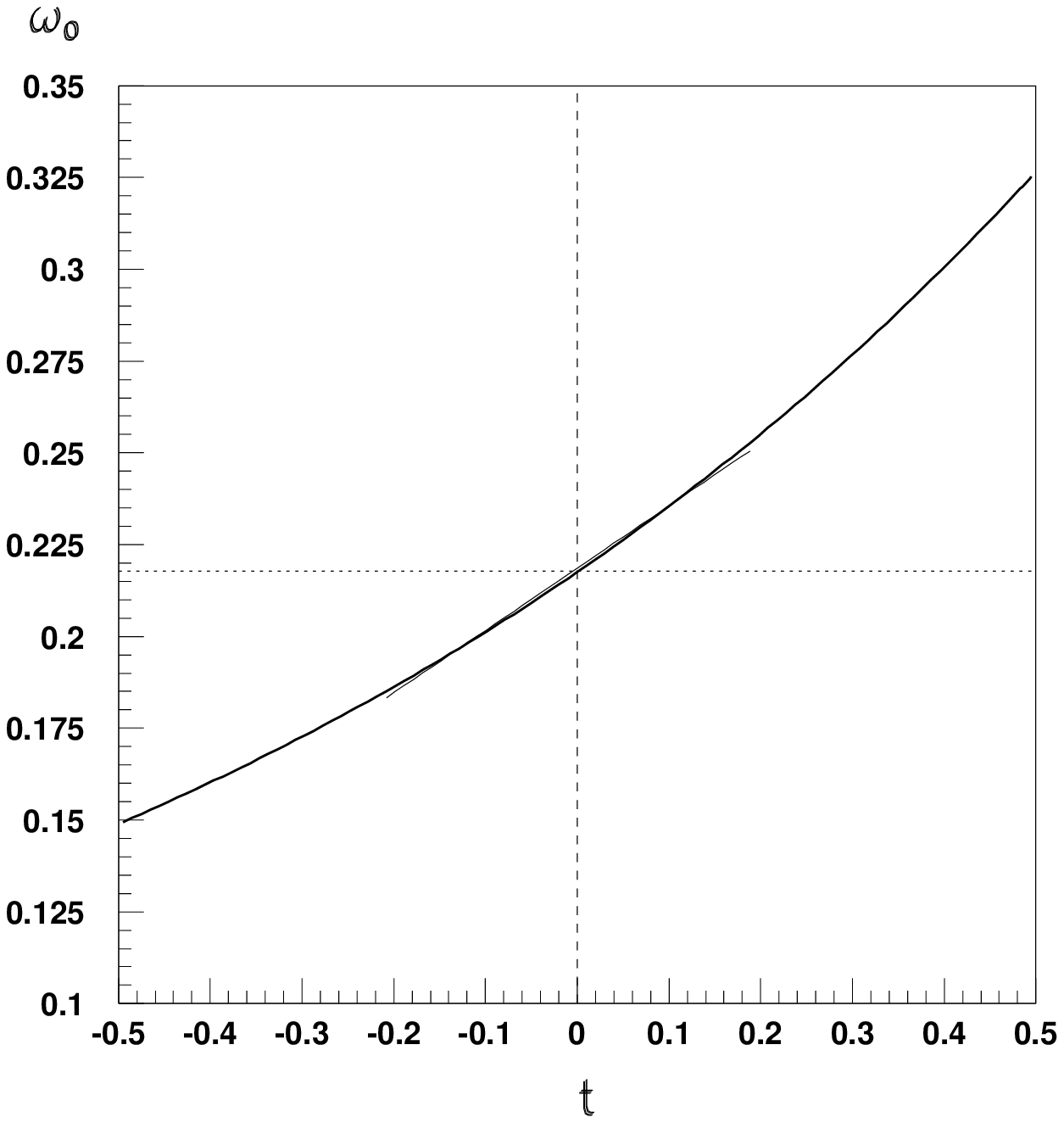,width=12cm}
\caption{The behaviour of the pomeron trajectory in the vicinity of $t=0$. }
\end{center}
\end{figure}

 We would like to mention that the Pomeron trajectory has about
the same slope ( $\alpha'_P(0)$ ) as the gluon trajectory.
More
interesting would be, within the same approach, to compare 
the Pomeron trajectory with the Reggeon trajectory.  
In order to do so one has to calculate the Reggeon trajectory 
in 2 + 1 QCD using the techniques developed in Ref. \cite{KL}. It 
will be a challenging problem for the future. 

At larger $|t|$ the trajectory deviates strongly from the linear behaviour.
It goes to infinity for $t$ approaching the threshold $t=4$
and returns from $-\infty$ above the threshold.
The behaviour of the Pomeron trajectory near $t=4$ has been obtained
also by solving Eq.(2.13) in the asymptotics of large $\omega$
and $t \rightarrow 4$ with the result
\beq
\omega_0 (t)\vert_{t \rightarrow 4} =
{g^2 \over 2 \pi m } \ \ \  {A_0 \over 4 - t }.
\label{t=4}
\eeq
This confirms the numerical result of Fig.~2.

The branch points $\om_1=0$, $\om_2=-\frac{g^2}{\pi m}$ do not depend on
$t$. However the singularities located at the unphysical sheet can come up
to the physical sheet as $t$ increases. Indeed both the poles arising from
the lower coefficients $\be_i$ in the hypergeometric functions and the
poles at the vanishing determinant depend on $q$. There are also branch
points arising from the square roots in the expressions of $\la^\pm$ in
terms of $\om$ (\ref{2.2.3}) the position of which depend on $q$.
The numerical investigation of the solution shows that besides of the pole,
which was originally the leading one, another pole emerges from
the unphysical sheet if $t$ crosses the threshold value.

Now we investigate the behaviour at $t \to -\infty$.
In the limit of large $ q$ we have
$$
\lambda^\pm = k\mp \frac{iq}{2}+O(\frac{1}{q}) \ ,
\ k=\sqrt{\frac{4+5\epsilon}{1+2\epsilon} } \ ,
$$

and

$$
d^b_a=a \frac{iq}{2} +\tau^{(a \cdot b)} \ ,
\ \tau^{(a)}=\frac{ a\sqrt{3} -1}{2} \ .
$$

Inserting these relations into the linear system we have found
that asymptotically in $q$

$$
f_0=-\frac{1}{2q^2}
$$

$$
A_1^+= -A_1^-
$$

$$
A_1^+ =iqf_0 \left\{
\frac{2f_2}{k+1}+\frac{2f_4}{k-1}-\frac{f_1}{k+2}-\frac{f_3}{k-2}
\right\}^{-1}
$$

where

$$
f_1={}_3F_2(^{k+\tau^{(+)},k+\tau^{(-)},k+2}_{2k+1,k+1}|y)
$$
$$
f_2={}_3F_2(^{k+\tau^{(+)},k+\tau^{(-)},k+3}_{2k+1,k+2}|y)
$$
$$
f_3={}_4F_3(^{k+\tau^{(+)},k+\tau^{(-)},k+3,k-2}_{2k+1,k+1,k-1}|y)
$$
$$
f_4={}_4F_3(^{k+\tau^{(+)},k+\tau^{(-)},k+3,k-1}_{2k+1,k+1,k}|y) \;\;.
$$

As the result we obtain from Eq.(\ref{solution-partial}) the asymptotics of
the partial wave

$$
F_\omega (q^2)|_{q^2\to \infty} \to \frac{\epsilon}{q^2}
\left\{
1-
\frac{2f_2}{k+1}/
(\frac{2f_2}{k+1}+\frac{2f_4}{k-1}-\frac{f_1}{k+2}-\frac{f_3}{k-2})
\right\}
$$

The behaviour of $F_\omega $ near the right branch point $\omega =0$ is
$F_\omega|_{q^2 \to \infty} \sim \frac{const}{\omega}$. This is to be
compared with the behaviour at the same point for $q=0$,
$F_\omega|_{q^2=0} \sim const\;\;.$
 The numerical  calculation of the Pomeron  trajectory
( see Fig.~2 )   shows that the pole is moving towards the right branch point
with decreasing $t$. From both observations we conclude that the Pomeron
pole moving with $t$ reaches the right branch point $\om=0$
asymptotically for $t=-q^2\to -\infty$.

\section{Comparison with the massless case}
\setcounter{equation}{0}

M.~Li and C.I.~Tan \cite{TAN} investigated 2 + 1 dimensional QCD
without symmetry breaking, i.e. for massless gluons, and obtained
just a fixed cut starting at $j=1$ as the leading singularity in
the vacuum exchange channel.
We try to understand the relation of their result to ours, in particular
whether the pomeron pole is absent in the massless case and how it
disappears at $m \to 0$.

The infrared singularities in 2 + 1 dimensions are stronger compared to
3 + 1 dimensions. The limit $m \to 0$ has to be performed carefully.
Clearly, the scattering amplitude
of vector bosons has no finite limit at $m \to 0$. We consider
the scattering of two colour dipoles of transverse sizes $x_1$, $x_2$, which
is the case studied in \cite{TAN}. The partial wave of the dipole--dipole
forward scattering is given by the convolution of two dipole impact factors
\cite{TAN} (here $x_0$ is the size of the dipole )
\beq
\Phi_D(x,k) = A\sin^2 k x_0
\label{dipol-if}
\eeq
with the Reggeon Green function
\beq
F^D_\om (x_{10},x_{20}) = \int \frac{d\,k_1\;\;d\,k_2}{(2\pi)^2}\;
\frac{\Phi_D(x_{10},k_1)}{k_1^2 + m^2}\;G_\om (k_1,k_2)\;
\frac{\Phi_D(x_{20},k_2)}{k_2^2 + m^2} \;\;.
\label{dipol-gf}
\eeq
The Reggeon Green function is the particular solution of the BFKL
equation with $\de$-functions as inhomogeneous term. It is related to
our solution $f_\om(k)$ which is more closely related to the vector
boson scattering as follows
\beq
\frac{A_0}{\om\,(k^2 + m^2)}\;f_\om(k)
= \int\limits^\infty_{-\infty}\;\;    G_\om (k,k_1)
\frac{d\,k_1}{k_1^2 + m^2}\;\;.
\label{dipol-scatt}
\eeq
Near the pomeron pole we have
\beq
G_\om (k_1,k_2) \approx \frac{\psi_0(k_1) \, \psi_0(k_2)}{\om - \om_0}
\;,\;\;\;  \om_0 = \frac{g^2}{m}\,\frac{1}{2\pi \epsilon_0} \;\;,
 \label{dipol-pom}
\eeq
where $\psi_0(k)$ is the wave function of the two-boson
bound state corresponding to the pomeron. It is normalized to 1 and can be
obtained from $f_\om$ by studying (\ref{dipol-scatt}) near $\om_0$.
$\epsilon_0$ is the number quoted in Eq.(3.32).

Restoring the mass dependence we obtain from the solution
(\ref{2.1.7})
\beq
f_\om(k) = \frac{1}{m^2}\;\phi (\frac{g^2}{m}\,\frac{1}{2\pi \om}, \frac{k}{m})
\label{dipol-mass}
\eeq
The solution depends smoothly on $k$ and the integral with a bounded
function $\Phi_D(x,k)$ exists. Therefore also
$\psi_0(k)$ has these features:
\beq
\psi_0(k) = \frac{a}{\sqrt{m}} \tilde{\phi}(\frac{k}{m}) \;\;,
\label{dipol-psi}
 \eeq
with $a$ being some numerical constant.
The contribution of the Pomeron pole to the scattering of two colourless
dipoles with sizes $x_{10}$  and $x_{20}$
  is given at small $m$ by the partial wave
\beq
F^D_\om \approx \frac{b\,\,g^4}{\om -
\frac{g^2}{m}\frac{1}{2\pi\epsilon_0}}\,m\,x_{10}^2\,x_{20}^2 \;\;.
\label{dipol-pw}
\eeq
Here $b$ is some number.
This leading contribution to the forward scattering of of dipoles does not
behave smoothly at $m\to 0$. The pole goes to plus infinity, resulting in a
divergent contribution. Expanding in $g^2$ we observe that the divergence
starts at the $g^4$
term, corresponding to $s$-channel intermediate state with two additional
gluons.

This observation is confirmed by calculating $G_{\omega} (k_1,k_2) $ 
iteratively and evaluating the corresponding contribution to the dipole scattering partial wave 
Eq.(5.2) in the following way.
We have to iterate Eq.(2.13) with the inhomogeneous term  replaced by 
$\delta (k_1 - k_2)$, which is the zeroth approximation of $G_{\omega}$. 
Unlike above in sect. 3.2 and 4.1 the iteration now proceeds order by order 
in $g^2$ or $\epsilon$. Replacing $G_{\omega}$ in (5.2) by $
\delta (k_1 -k_2) $  we obtain that the region of $k_1, k_2 \sim m $ gives 
a negligible contribution for $m \rightarrow 0 $. Taking the first order 
approximation in $\epsilon$ for $G_{\omega}$ leads to a finite contribution 
of that small- $k$  region. With the ${\cal O} (\epsilon^2 )$ approximation 
for $G_{\omega}$ we obtain a like ${1 \over m}$ diverging contribution.
 Starting from this order of perturbative expansion the amplitude of
forward dipole - dipole scattering does not exist in the massless limit.

Consider now the scattering at non-vanishing momentum transfer. Let us
fix the value $t_{phys}$ in physical units (GeV$^2$) and look at the relation
to our dimensionless
variable $t=-q^2$ (in units of $m^2$)
\beq
t_{phys}=t\;m^2  \;\;.
\label{dipol-t}
\eeq
Provided $t_{phys}<0$, the corresponding value of $t$ approaches
$-\infty$ at $m\to 0$. Thus the pomeron pole approaches the
 branch point at $j=1$.

The singular contribution (\ref{dipol-pw}) appearing only at
$t_{phys}=0$ is absent in the infrared finite dipole scattering amplitude
constructed in \cite{TAN}.

Let us now study the massless limit directly in the equation. We restore
the masses in Eq.~(\ref{2.5}) and do the shift $k \to k-\frac{q}{2}$ as in
Eq.(\ref{2.2.2})
\beqar{bfkl-shift}
&&\left[
1+ \epsilon ( \frac{(k-\frac{q}{2})^2+m^2}{(k-\frac{q}{2})^2+4m^2}+
\frac{(k+\frac{q}{2})^2+m^2}{(k+\frac{q}{2})^2+4m^2} )
\right] f_\omega (k,q) = \nonumber \\
&& = A_0^{-1}+
\epsilon m \int \frac{\displaystyle d k_1}{2\pi }\times
\left(
\frac{A_0}{\displaystyle ((k_1 +\frac{q}{2} )^2+m^2)((k_1-\frac{q}{2})^2+m^2)}
\right. \nonumber \\
&& \left. +
\frac{2}{(k-k_1)^2+m^2}
\left[
\frac{(k+\frac{q}{2})^2+m^2}{(k_1 + \frac{q}{2})^2+m^2}+
\frac{(k-\frac{q}{2})^2+m^2}{(k_1-\frac{q}{2})^2+m^2}
\right]
\right) f_\omega (k_1,q) \;\;.
\eeqar

We perform the Fourier transformation with respect to $k$
\beq
\label{fourier}
f_\omega (x,q) = \int\;\frac{dk}{2\pi} e^{-ikx} f_\omega (k,q)
\eeq
and obtain
\beqar{gen-eq}
&&\left(1 + 2\epsilon  -2\epsilon\, e^{-m|x|}  \right)f_\omega (x,q) -
\frac{3}{2}\epsilon m
\int dy\, f_\omega (y,q)\, \cos \frac{q }{2}(y-x) \,e^{-2m|x-y|} = \nonumber \\
&& = \frac{1}{A_0}\de(x)
+ \epsilon \de(x)\left(\frac{A_0}{q^2 + 4m^2} +2\right)
\int dy \,f_\omega (y,q)\,e^{-m|y|}\,\cos \frac{q }{2}y \nonumber \\
&& + \frac{2A_0\,\epsilon m \,\de(x)}{q(q^2 + 4m^2)}
\int dy \,f_\omega (y,q)\,e^{-m|y|}\,\sin \frac{q }{2}|y|  \\
&& -\epsilon m \,e^{-m|x|}\int dy \,f_\omega (y,q)
\,\cos \frac{q }{2}(y-x)\, e^{-m|x-y|}\,(2\,\mbox{sgn}(x)\,\mbox{sgn}(x-y) +1)\;\;. \nonumber
\eeqar
 It should be stressed that \eq{gen-eq} is the general BFKL equation
for 2 + 1 QCD in the coordinate space at any value of the momentum
transfer $ t = - q^2$.

As discussed above the behaviour at $m=0$ is different for forward
and non-forward cases. Indeed the coefficient of second term on r.h.s.
$\epsilon \left( \frac{A_0}{q^2 + 4m^2} + 2 \right)$ vanishes at
$m\to 0$ for $q \neq 0$ but behaves like $\frac{1}{m}$
if we put $q=0$ before taking the limit $m\to 0$. We discuss in the following
the massless limit in the non-forward case. We approximate Eq.~(\ref{gen-eq})
at $m\to 0$ expanding in particular $e^{-m|x|}$ keeping terms
$\epsilon m$ (because $\epsilon = \frac{g^2}{2\pi m}$). In this way we obtain
\beqar{massless}
&&\left(1 + 2\epsilon m|x|\right) {f}_\om(x,q) -\frac{1}{2}\epsilon m
\int dy \,{f}_\om(y,q) \,\cos \frac{q}{2}(x-y) = \nonumber \\
&&\frac{1}{A_0}\,\de(x) - 4\de(x)\,\frac{\epsilon m}{q}
\int dy \,{f}_\om(y,q) \,\sin \frac{q}{2}|y| \nonumber \\
&& -2\,\epsilon m \,\mbox{sgn}(x)\int dy \,{f}_\om(y,q)\,\cos \frac{q}{2}(x-y)
\,\mbox{sgn}(x-y) \;\;.
\eeqar
In terms of the function $\tilde{f}_\om(x,q)$ defined as
\beq
\label{f-tilde}
{f}_\om(x,q)  = - \left( \frac{d^2}{dx^2} + \frac{q^2}{4}\right)
\tilde{f}_\om(x,q)
\eeq
Eq, (\ref{massless}) has the following simple form
\beq
\label{leq}
\left(\frac{d^2}{dx^2} + \frac{q^2}{4}  \right)
\left[(1 + 2\epsilon m|x|)\tilde{f}_\om(x,q) \right] = -\frac{\de(x)}{A_0}\;\;.
\eeq

The solution has the form
\beqar{solution}
&&(1 + 2\epsilon m|x|)\tilde{f}_\om(x,q) = \nonumber \\
&& = -\frac{1}{qA_0}\;\sin \frac{q}{2}|x| + A \sin \frac{q}{2}x
+ B \cos \frac{q}{2}x
\eeqar
where $A, B$ are some arbitrary constants.

The original function $f_\om(x,q)$
can be calculated readily. We write here only
part of the result with $A=B=0$, which corresponds to certain boundary
conditions
\beqar{solAB}
&&f_\om(x,q) = \nonumber \\
&&= \frac{1}{A_0}\left[  \frac{\de(x)}{1+2\,\epsilon m\,|x|}
- \frac{2\epsilon m\cos \frac{q}{2}|x|}{(1+2\epsilon m\,|x|)^2}
+ \frac{8(\epsilon m)^2\sin \frac{q}{2}|x|}{q(1+2\epsilon m|x|)^3}
         \right] \;\;.
\eeqar
This result has similarities to the expression for the dipole density
$n_\om(x_0,x,q)$ derived by Li and Tan (Eq.(3.4) in the second paper of
Ref. \cite{TAN}). Our $f_\om(x,q)$ is not the dipole density and
 our equation does not know about the dipole size on which $n_\om(x_0,x,q)$
depends essentially. However, introducing the dipole size $x_0>0$
by replacing the r.h.s. of (\ref{leq}) by $-\de(x_0 - |x|)$ and
restricting the range in $x$ to $|x| \neq 0$, we
obtain
\beqar{green}
&&G_\om(x,x_0;q) = (\frac{d^2}{dx^2} + \frac{q^2}{4})\;
\frac{1}{[1+2\epsilon m|x|]}
\;\frac{2}{q}\;\theta(x_0 - |x|)\,\sin \frac{q}{2}(x_0-|x|) \nonumber \\
&&= \frac{\de(x_0 -|x|)}{1+2\,\epsilon m\,|x|}
+ \theta(x_0-|x|)
\left[\frac{16(\epsilon m)^2\sin \frac{q}{2}(x_0-|x|)}{q(1+2\epsilon m|x|)^3}
\right. \nonumber \\
&& \left. + \frac{4\epsilon m\cos \frac{q}{2}(x_0-|x|)}{(1+2\epsilon m\,|x|)^2}
\right]  \;\;.
\eeqar
We denote by $G_\om(x,x_0;q)$ the analogon of $f_\om(x,q)$ (\ref{f-tilde})
of the modified equation. This particular solution of the modified
Eq.~(\ref{leq}) reproduces the dipole density
$n_\om(x_0,x,q)$ of Ref. \cite{TAN} up to  terms proportional to
$\de(x)$.

%%%%%%%%%%%%%%%%%%%%%%%%%%%%%%%%%%%%%%%%%%%%%%%%%%%%%%%%%%%%%%%%%%%%%%%

%%%%%%%%%%%%%%%%%%%%%%%%%%%%%%%%%%%%%%%%%%%%%%%%%%%%

 \section{Summary}
\setcounter{equation}{0}

%%%%%%%%%%%%%%%%%%%%%%%%%%%%%%%%%%%%%%%%%%%%%%%%%%%

The reduction of the dimensionality to 2 + 1 simplifies the high-energy
scattering amplitudes and in particular the BFKL equation. The equation 
can be solved  analytically  even in the case with masses introduced 
by spontaneous symmetry breaking.

In the forward case we have discussed the solution both in coordinate 
and in momentum space. In the coordinate space the similarity of the
Pomeron pole to a two gauge boson bound state has been emphasized,
whereas in the momentum representation the iterative structure becomes
 transparent, which has been used further to solve the equation in the
non-forward case.

We obtain the partial wave for the scattering amplitude of vector 
bosons in an analytic form. This allows us to study the leading and 
non-leading Regge singularities and their dependence on the momentum
transfer both for negative and positive $t$.

At small momentum transfer we find a leading Regge pole, the Pomeron,
with an intercept above $j=1$ and a trajectory approximately linear
for small $t$. Beside this pole there is a branch cut whose right end
is located at $j=1$  independent  of $t$.
The Pomeron pole approaches the cut for large negative $t$.

We have compared our result with the one by M.~Li and C.-I.~Tan \cite{TAN}
and have discussed the peculiarities of the massless limit.
This limit is different for the forward and non-forward case.
The massless limit of our result for the non-forward amplitude 
is close to the result found
by Li and Tan  who used a quite different approach.
 A modified equation 
(by introducing the dipole size as an additional parameter) 
reproduces the result of these authors. However, in the forward case
the Pomeron pole leads to a
divergent contribution, which is absent in the
amplitude of Ref.~\cite{TAN}.

Although the model lives in unphysical 2 + 1 dimensions some features 
can be related to the phenomenology of high-energy scattering.
The leading pomeron pole which is clearly separated from non-leading
singularities corresponds to the common idea about the soft
(non-perturbative) pomeron. Furthermore, a situation with a pole with 
intercept larger than 1 and a fixed cut just at 1 would result in a
change of the $s$-dependence with $t$ . {It is interesting to note 
that the constant term in the high energy asymptotics, which correponds to 
the cut contribution, has been used to
 describe the experimental data on  high energy behaviour of
 total cross sections  \cite{TOTS},
inclusive spectra \cite{INCS} and diffractive dissociation 
\cite{TANF}.}

The divergence of the trajectory at $t \to 4m^2$ seems to exhibit an infinite
series of bound states of two massive gluons. This would 
differ clearly from the features of the hadronic reality.
We understand that this is an artifact of the leading $\ln s$ 
approximation, since a potential of finite range created by massive boson 
exchange cannot have an infinity of bound states.

The simplicity of the model makes it useful for further investigations.
Including fermions the amplitudes with quantum number exchange could be
constructed. There will be no direct analogy neither to DGLAP 
\cite{DGLAP} equation nor to
the nonlinear double-log equation \cite{KL}. More interesting could 
be the study 
of amplitudes with multiple exchange of reggeized gluons, in particular
with the exchange of negative charge parity (Odderon). 

The model can serve as a
testing ground for the non-perturbative treatment of diffractive processes
\cite{SCGLUON1}, \cite{SCGLUON2}, and the high parton density effective action
\cite{MCLER}. Also, the effective action of high-energy
scattering \cite{KLS} and Gribov's reggeon field theory
 \cite{GRIB}
can be studied in the simpler situation of 2~+~1~dimensions.

\vspace*{1cm}

{\Large \bf  Acknowledgments:}

\centerline{}

E.M.L.,L.N.L. and M.W. want to thank the Aspen Physics Centre for the
hospitality and creating the working atmosphere during the workshop on "
Interface of ``soft" and ``hard" interactions" in July 1996, where this
paper has been started. We are also very grateful to C-I Tan who draw
our attention to the problem of the mismatch between massive and massless 
2 + 1 dimensional QCD.
Without his questions and his strong encouragement this paper
would not have been started at all. 
E.M.L. thanks the theory groups at Fermilab,
ANL and DESY where he continued to work on this paper for hospitality and
support. 

D.Yu I. and L.Sz. would like to acknowledge the warm hospitality 
extended to them at University of Leipzig.

\end{document}